\documentstyle[12pt,epsf]{article}
\newcommand{\zum}  {{\rm{\Sigma}}}
\newcommand{\gs}   {\gamma_{\rm S}}
\newcommand{\gsi}  {\gamma_{{\rm S}i}}
\newcommand{\gsz}  {\gamma_{{\rm S}0}}
\newcommand{\QG}   {{\bf{Q}}}
\newcommand{\QGz}  {{\bf{Q}}^0}
\newcommand{\QGzi} {{\bf{Q}}_i^0}
\newcommand{\QGi}  {{\bf{Q}}_i}
\newcommand{\LS}   {\lambda_{\rm S}}
\newcommand{\ppb}  {$\rm{p+\bar{p}}\;$}
\newcommand{\pp}   {$\rm{p+p}$ }
\newcommand{\nn}   {$\rm{n+n}$ }
\newcommand{\pn}   {$\rm{p+n}$ }
\newcommand{\ee}   {${\rm e}^++{\rm e}^-$}

\newcommand{\qj}   {{\bf{q}}_j}

\newcommand{\dint} {{\rm{d}}}
\newcommand{\E}    {{\rm{e}}}

\newcommand{\muv}  {{\boldmath $\mu\!$ \unboldmath}}
\newcommand{\muvs} {{\mbox{\boldmath ${\scriptstyle \mu}\!$ \unboldmath}}}

\newcommand{\oV}   {{\overline V}}

\newcommand{\oB}   {{\overline B}}
\newcommand{\oQ}   {{\overline Q}}

\newcommand{\vx}   {{\bf{X}}}
\newcommand{\va}   {{\bf{a}}}
\newcommand{\ovx}  {{\overline \vx}}

\begin{document}
\begin{flushright}
DFF 292/10/1997 \\
IKF--HENPG/4--97 \\
BI--TP 97/42
\end{flushright}
\vspace{0.5cm}

\begin{center}

{\Large \bf On Chemical Equilibrium \\[5mm]
in Nuclear Collisions}

\vspace{0.8cm}

F. Becattini\footnote{E--mail: becattini@fi.infn.it}\\
INFN Sezione di Firenze\\
Largo E. Fermi 2, I - 50125 Firenze, Italy\\[0.5cm]

M. Ga\'zdzicki\footnote{E--mail: marek@ikf.uni--frankfurt.de}\\
Institut f\"ur Kernphysik, Universit\"at Frankfurt \\
August--Euler--Strasse 6, D - 60486 Frankfurt, Germany\\[0.5cm]

J. Sollfrank\footnote{E--mail: sollfran@physik.uni-bielefeld.de}\\
Fakult\"at f\"ur Physik, Universit\"at Bielefeld\\
Universit\"atsstr. 25, D - 33615 Bielefeld, Germany\\

\vspace{2cm}
\begin{center}
{\bf Abstract}
\end{center}
\vspace{1cm}

\begin{minipage}{14cm}
\baselineskip=12pt
\parindent=0.5cm

{\small 
The data on average hadron multiplicities in central A+A collisions 
measured at CERN SPS are analysed with the ideal hadron gas model.
It is shown that the full chemical equilibrium version of the model
fails to describe the experimental results. The agreement of the
data with the off--equilibrium version allowing for partial
strangeness saturation is significantly better. The freeze--out 
temperature of about 180 MeV seems to be independent of the system 
size (from S+S to Pb+Pb) and in agreement with that extracted in 
\ee, \pp and \ppb collisions. The strangeness suppression is discussed 
at both hadron and valence quark level. It is found that the
hadronic strangeness saturation factor $\gamma_{\rm S}$ increases 
from about 0.45 for \pp interactions to about 0.7 for central A+A 
collisions with no significant change from S+S to Pb+Pb collisions.
The quark strangeness suppression factor $\LS$ is found 
to be about 0.2 for elementary collisions and about 0.4 for heavy ion
collisions independently of collision energy and type of colliding 
system.}

\end{minipage}

\vspace{1cm}
{\it Submitted to Z. Phys. C}
\end{center}

\newpage

\section{Introduction}

Many years of experimental effort in the field of high energy nuclear
collisions yielded a large amount of data on particle production
at different collision energies (up to 200 $A$ GeV$/c$) and for different
colliding systems \cite{QM96}. These results allow studying the properties
of strongly interacting matter at high energy densities. Ultimately, 
at high enough collision energy one expects to create in the laboratory
the Quark--Gluon Plasma (QGP), a form of matter in which effective 
degrees of freedom are quarks and gluons instead of hadrons and hadronic 
resonances. The formation of QGP with deconfinement of quarks and
gluons should hopefully be reflected in the final hadron production,
provided that the expected modifications of the entropy and strangeness 
content of the system survive hadronization and reinteractions between final
state hadrons. In principle the system evolution is determined by QCD.
Nevertheless, the formation of hadrons is a process entirely lying in
its non--perturbative domain, hence, in order to study the final
state, one has to resort to phenomenological models such as string 
or statistical (thermal) models.

The statistical models, whose prototypes \cite{Fermi50,Landau53,Hagedorn60}
date back to '50s and '60s, are based on the assumption of local filling
of available phase space according to statistical laws, once collective
effects have been taken into account. This {\it ansatz} allows the
characterization of hadron production by means of few parameters such as
temperature, volume and chemical potentials. Furthermore, parameters 
accounting for possible departures from complete equilibrium are often 
introduced.

A long and rich history of thermal models is related to their surprising 
success in the description of many aspects of high energy collisions
\cite{Divonne}.
It has been shown recently that the ideal hadron gas model allowing for 
non--equilibrium strangeness abundance is able to reproduce hadron 
multiplicities in \ee, \pp and \ppb interactions over a large collision 
energy range \cite{Becattini96a,Becattini96b,Becattini97a}.
In this paper we apply the same model to analyse the data of hadron 
multiplicities in central nucleus--nucleus collisions at CERN SPS collision 
energies. The data on Sulphur--nucleus collisions at 200 $A$ GeV$/c$ 
were already analysed using thermal models in many previous works (for 
a review see Ref.~\cite{Sollfrank97b}). We add to this data set the 
preliminary results on hadron production in central Pb+Pb collisions 
at 158 $A$ GeV$/c$. 
The use of the same formulation of the thermal model for 
the analysis of data from \pp to central Pb+Pb collisions
allows us to study the evolution of the parameters of the model
in the full range of the colliding systems. Our analysis is based on 
hadron multiplicities integrated over the full phase space because
their use minimizes the influence of collective effects on final results 
allowing a formulation of the model to be tested with a minimal 
number of assumptions and parameters.  

The paper is organized as follows: in Sect.~2 we describe the hadron 
gas model which is confronted with the experimental data in Sect.~3.
Discussion and conclusions are given in Sect.~4.

\section{Hadron Gas Model}

In this section the hadron gas model used in the present analysis is 
sketched; a more detailed description can be found in 
Refs.~\cite{Becattini96b,Becattini97a}.
The model postulates the formation of an arbitrary number of hadron gas 
fireballs each having a definite collective momentum as the result 
of an interaction between the two colliding systems. The parameters 
describing a hadron gas fireball at thermal and chemical equilibrium are the 
temperature $T_i$, the volume $V_i$ in its rest frame, and the quantum 
numbers, i.e. electric charge $Q_i$, baryon number $B_i$ and strangeness $S_i$. 
Charm and beauty have been excluded from the calculations as the thermal 
production of heavy--flavoured hadrons is negligible with respect to 
non--heavy-flavoured hadrons in the expected range of temperatures, 
namely 100$\div$200 MeV. If $\QGzi=(Q_i,B_i,S_i)$ is the vector of $i^{th}$ 
fireball's quantum numbers and $N$ is the number of fireballs, the 
following constraint:

\begin{equation}
  \sum_{i=1}^N \QGzi = \QGz \; ,
\end{equation} 
where $\QGz$ is the quantum vector fixed by the initial state, must be
fulfilled. 
The average yield of any hadron in the $i^{th}$ fireball can be derived 
from the partition function:

\begin{equation}
  Z_i(\QGzi) = \sum_{\rm{states}} \E^{-E_i/T_i} \delta_{\QGi,\QGzi} \; ,
\end{equation}
which is calculated in the canonical approach, i.e. by using only the 
multi--hadronic states having the same quantum numbers of the fireball.

A non--equilibrium parameter $\gsi$ accounting for a possibly incomplete 
strangeness chemical equilibration is introduced by multiplying by 
$\gsi^s$ the Boltzmann factors $\E^{-E_j/T_i}$ associated 
to the $j^{th}$ hadron in the partition function, where $s$ is the number 
of its valence strange quarks and antiquarks. Although this factor
was introduced heuristically \cite{Rafelski91} and used as a purely 
phenomenological parameter in the analysis of elementary collisions
\cite{Becattini96a,Becattini96b,Becattini97a}, it can be shown that 
$\gs$ formally is the fugacity associated to the number of 
strange + antistrange quarks in the hadron phase in a grand-canonical 
framework \cite{Slotta95}. 

The overall average multiplicity of each hadron species is the sum 
of all average yields in each fireball. In principle any configuration
$\{\QG_1^0,\ldots,\QG_N^0\}$ of fireballs in the event may occur, so that 
average hadron abundances depend on the probability $w(\QG_1^0,\ldots,\QG_N^0)$
of occurrence of a given configuration besides the whole set of fireball 
thermal parameters $T_i$, $V_i$ and $\gsi$. 
However, it can be shown that if such configuration weights 
$w(\QG_1^0,\ldots,\QG_N^0)$ are chosen in a statistical fashion 
\cite{Becattini96b,Becattini97a}:

\begin{equation} \label{weight}
 w(\QG_1^0,\ldots,\QG_N^0) = \frac{\delta_{\zum_i \QGzi,\QGz} \prod_{i=1}^N 
 Z_i(\QGzi)}{\sum_{\QG_1^0,\ldots,\QG_N^0} \!\!\! \delta_{\zum_i \QGzi,\QGz} 
 \prod_{i=1}^N Z_i(\QGzi)}  
\end{equation}
and the fireball freeze--out temperatures and $\gsi$ suppression 
factors are the same, namely $T_i\equiv T$ and $\gsi \equiv \gs$, then the 
average hadron abundances $n_j$ at freeze--out depend only on the {\it global 
volume} $V\equiv \sum_{i=1}^N V_i$ and $\QGz$ through the following equation:
\begin{equation}  \label{three}
 n_j = (2J_j+1)\, \frac{VT}{2\pi^2} \sum_{l=1}^{\infty} (\mp 1)^{l+1} \, 
 \frac{Z(\QGz-l\qj)}{Z(\QGz)} \,\gs^{l s_j} \frac{m^2_j}{l} 
 {\rm K}_2(\frac{lm_j}{T}) \; ,
\end{equation}
where the upper sign is for fermions and the lower for bosons; the function 
$Z$ is the {\it global partition function} \cite{Becattini96b,Becattini97a} 
and $\qj$ is the quantum number vector of the $j^{th}$ hadron species. 

The special choice of weights $w(\QG_1^0,\ldots,\QG_N^0)$ in 
Refs.~\cite{Becattini96b,Becattini97a} leads to the same expression of 
average 
multiplicities relevant to a system in global equilibrium even if
the fireballs are not in mechanical equilibrium. It might be
argued that the difference between the rapidity spectra of baryons and 
antibaryons -- existing even in central A+A collisions \cite{Al:94} --
question the validity of this choice. Nevertheless this choice has a
remarkable property, namely it removes the dependence of hadron average 
multiplicities on both the number of fireballs and their ordering 
either in size or in space (reabsorbed in the global volume). As a 
consequence, much freedom is left to possibly reproduce the rapidity 
distributions within the model keeping the same quantitative expressions 
for hadron abundances (see Appendix A). 
However, even if the choice of the weights $w(\QG_1^0,\ldots,\QG_N^0)$ was 
not correct, the corrections to eq.~(\ref{three}) are expected to be 
small because the relative abundances are predominantly determined by 
the intensive thermal parameters and not by fireball quantum configuration 
weighting. 

The {\it chemical factors} $Z(\QGz-l\qj)/Z(\QGz)$ in eq.~(\ref{three})
implement the dependence of the yields upon the chemistry of the system 
and replace the chemical potentials in the proper canonical formalism. 
For very large volumes, as expected in heavy ion collisions, it can be 
proved \cite{Becattini97a} by means of a saddle--point approximation, that 
eq.~(\ref{three}) reduces to:

\begin{equation} \label{four}
  n_j = (2J_j+1)\, \frac{VT}{2\pi^2} \sum_{l=1}^{\infty} (\mp 1)^{l+1} \, 
 \gs^{l s_j} \frac{m^2_j}{l} {\rm K}_2(\frac{lm_j}{T}) \,\, 
   \E^{l \QGz {\sf{A}}^{-1} \qj/2} \,\, \E^{-l^2 \qj {\sf{A}}^{-1} \qj/4} \; ,
\end{equation}
where ${\sf{A}}$ is a $3\times 3$ matrix whose elements are proportional 
to $V$:

\begin{equation}\label{five}
 {\sf{A}}_{k,l} = \frac{1}{2} \sum_j \frac{V \, 
 (2J_j+1)}{(2\pi)^3} \int \dint^3 p \,\, \frac{\gs^{s_j} 
  \E^{-\sqrt{p^2+m_j^2}/T}}
 {(1\pm\gamma_{\rm S}^{s_j}\E^{-\sqrt{p^2+m_j^2}/T})^2} \,\, 
  q_{j,l} q_{j,k} \; ,
\end{equation}  
where the sum runs over all hadron species. In eq.~(\ref{four}) the 
chemical factors are transformed into a product of two factors:
the first one can be written as $\exp[l \mbox{\muv} \cdot \qj/T]$ 
where \muv is a traditional set of chemical potentials, whereas 
the factor $\exp[-l^2 \qj {\sf{A}}^{-1} \qj/4]$ has no grand-canonical 
corresponding quantity. Its presence entails a suppression of hadrons 
having non--vanishing quantum numbers with respect to neutral ones, owing to 
the finite size of the system. Indeed this factor takes its origin from 
the requirement of exact conservation of initial quantum numbers. In the 
infinite volume limit ${\sf{A}^{-1}}$ goes to zero and the grand-canonical 
formalism is fully recovered.

An important problem to face in modelling heavy ion collisions is the 
fact that particle multiplicities are measured by averaging over events
with a varying number of participant nucleons. For central collisions of 
identical nuclei (S+S and Pb+Pb in this paper) the fluctuations in the 
number of spectator nucleons, $N_{\rm SPEC}$, can be considered to follow 
a Poisson distribution \cite{Golokhvastov}, well approximated by a 
Gaussian distribution if $\langle N_{\rm SPEC} \rangle \gg 1$. 
Hence fluctuations in the number of participant nucleons, $N_{\rm PART} 
= 2 \cdot A - N_{\rm SPEC}$, can be described by a Gaussian
distribution with mean value $\langle N_{\rm PART} \rangle$ and variance 
$\langle N_{\rm SPEC} \rangle$. For the considered collisions 
(S+S and Pb+Pb\footnote{Validity of the above consideration for central 
S+Ag collisions is questionable as fluctuations are probably dominated 
by a geometrical effect.}) it is $\langle N_{\rm SPEC} \rangle \ll 
\langle N_{\rm PART} \rangle$, implying that the mean of participant
nucleons distribution is large in comparison with its width.
Assuming that $T$ and $\gs$ do not depend on the number of participant 
nucleons, the average number of directly produced hadrons of species 
$j$ is:

\begin{equation}\label{six}
 \langle n_j \rangle \!= \!(2J_j+1) \, \sum_{\QGz} \int \dint V  F(\QGz,V)
 \, \frac{VT}{2\pi^2} \sum_{l=1}^{\infty} (\mp 1)^{l+1} \, 
 \gs^{l s_j} \frac{m^2_j}{l} {\rm K}_2(\frac{lm_j}{T}) \,\, 
 \E^{l \muvs \cdot \qj/T} \,\, \E^{-l^2 \qj {\sf{A}}^{-1} \qj/4} \! ,
\end{equation}
where $F(\QGz,V)$ is the joint probability of observing an event with
global volume $V$ and quantum vector $\QGz$ (determined by the number 
of protons and neutrons involved in the collision). The removal of the 
integral from eq.~(6) by introducing mean chemical potentials 
$\overline{\mbox{\muv}}$ and mean volume $\oV$ must be undertaken with 
great care since such mean values in general would depend on the hadron 
species and would not be the same for all of them. This is clearly understood 
because \muv and $\sf{A}$, depending on $\QGz$ and $V$, multiply $\qj$ 
in a non--factorisable form.
However, if a reasonable {\it ansatz} of a narrow Gaussian shape for the 
distribution function $F(\QGz,V)$ is assumed, according to previous 
discussion, it can be shown (see Appendix B) that the simple averaging 
procedure introducing mean chemical potentials and volume can be used
provided that ${\sf{A}}^{-1} \ll 1$. In other words, for the following 
simple averaging formula to hold:

\begin{equation} \label{seven}
 \langle n_j \rangle \approx (2J_j+1)\, \frac{\oV T}{2\pi^2} 
 \sum_{l=1}^{\infty} (\mp 1)^{l+1} \, \gs^{l s_j} \frac{m^2_j}{l} 
 {\rm K}_2(\frac{lm_j}{T}) \, \E^{l \overline{\muvs} \cdot \qj/T} \,\, 
 \E^{-l^2 \qj {\overline{\sf{A}}^{-1}} \qj/4} \; ,
\end{equation}
a nearly grand-canonical regime and reasonably small fluctuations of
$V$ and $\QGz$ are needed. How small such fluctuations must be, will be 
checked {\it a posteriori} in the actually examined collisions. The hadron 
average multiplicities are calculated with eq.~(\ref{seven}) in which 
$\oV$, $T$, $\gs$ and $\overline{\mu}_B$ are free parameters to be 
fitted to the data, while $\overline{\mu}_S$ and $\overline{\mu}_Q$, the 
strangeness and the electrical chemical potentials respectively, have 
been determined by means of two additional constraints: 
the strangeness neutrality and the conservation of the ratio 
$(Z_1+Z_2)/(A_1+A_2)$ formed with the atomic and mass numbers of the two 
colliding nuclei:
\begin{eqnarray}
 && \sum_j S_j \langle  n_j \rangle = 0 \nonumber \\ 
 && \sum_j Q_j \langle  n_j \rangle = \frac{Z_1+Z_2}{A_1+A_2} 
    \sum_j B_j \langle  n_j \rangle  \; . \label{eight}
\end{eqnarray}
It should be noted that the first of the two constraints in 
eq.~(\ref{eight}) is valid on an event by event basis 
whereas the second, only averaged over a large number of collisions.

All previous equations are concerned with primary hadrons, namely
particles and resonances directly emitted from the hadronic source
and not coming from secondary decays. On the other hand, since actual 
measurements include feeding from heavier hadrons and resonances, the 
hadron production rates to be compared with the data 
have been calculated by letting all primary hadrons decay according to known
branching ratios until particles considered stable by the experiments 
are reached. Among primary hadrons we included all particles and resonances 
up to a mass of 1.7 GeV; the masses of resonances have been distributed 
according to a relativistic Breit--Wigner. The needed values of hadron 
masses, widths and branching ratios have been taken from the most recent 
Particle Data Book \cite{Barnett96}. 

\section{Comparison with the Data}
\subsection{Full Equilibrium Model}

The hadron gas model described in the previous section has been used to
fit the data on hadron abundances in central S+S, S+Ag at 200 $A$ GeV$/c$ and 
central Pb+Pb collisions at 158 $A$ GeV$/c$ measured by NA35 and NA49
Collaborations at CERN SPS. We used average hadron multiplicities measured 
in (or extrapolated to) full phase space. The compiled data and the 
references to the original papers, where the experimental details (acceptances, extrapolation
procedures) can be found, are given in Table~\ref{comparison}. Since 
preliminary data on central Pb+Pb collisions are still poor, we decided
in this case to use in addition two particle ratios measured in the central 
rapidity region, namely ${\rm K}^+/{\rm K}^-$ and $\overline{\Lambda}/\Lambda$.
The measured rapidity distributions of these particles in the acceptance 
region are similar to those of the corresponding antiparticles \cite{Bo:97}, 
thus justifying our decision to use those ratios as estimates of the
full phase space ones.

The analysis of particle multiplicities in \ee, \pp and \ppb collisions 
within a hadron gas model indicates that the strangeness production
in such elementary collisions is not high enough to ensure the complete
local chemical equilibrium at hadron level. It may be expected that in the case 
of central A+A collisions, due to a much larger volume of the interaction 
region and an increased role of hadron rescattering, a full local chemical
equilibration at hadronic level can be attained.

\begin{figure}
\caption[]{Particle ratios in the grand-canonical approximation for
$\gs =1$. The bands correspond to $\pm1\sigma$ deviations of the
experimental ratios summarized in Table~\ref{comparison}. The symbol
K stands for $\langle {\rm K} \rangle = \langle {\rm K}^+ \rangle + 
\langle {\rm K}^- \rangle + 2\langle {\rm K}^0_s \rangle$. 
\label{gammasone}} 

\begin{minipage}{7.9cm}
\epsfxsize 7.9cm \epsfbox{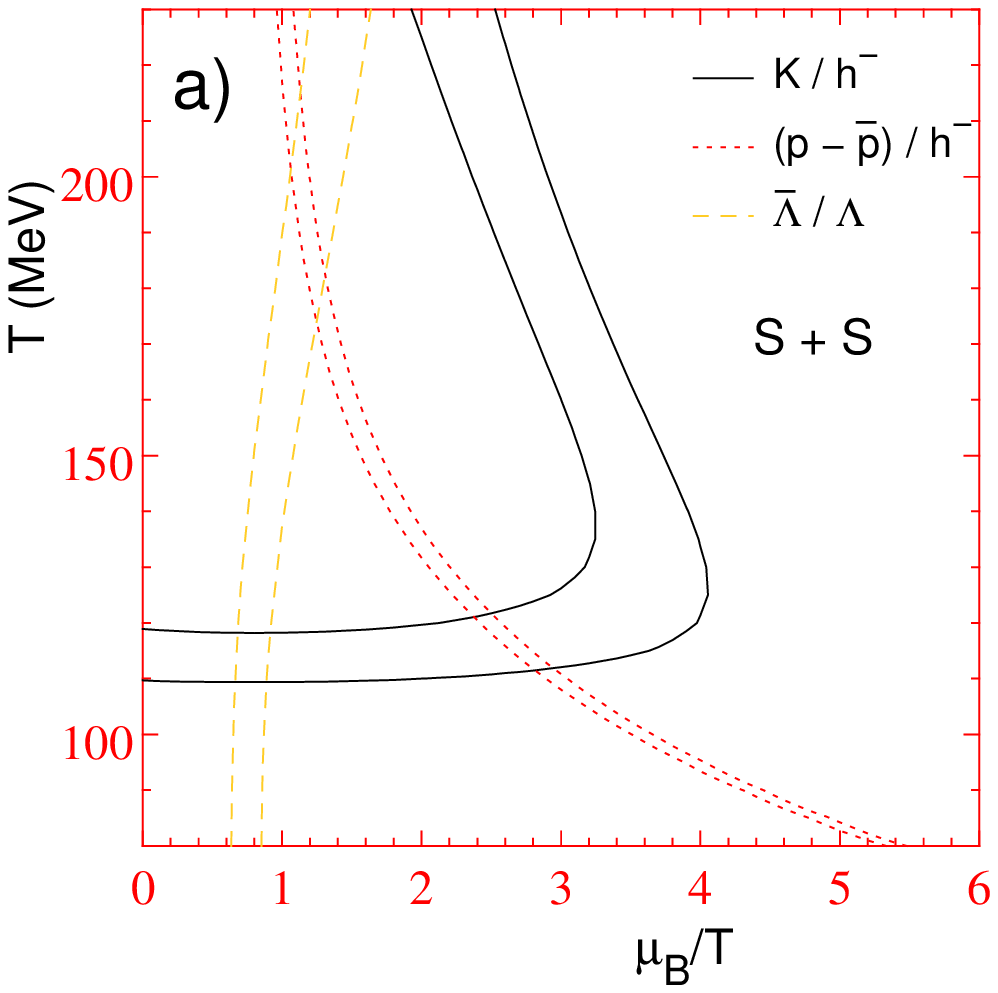}
\end{minipage}
\hfill
\begin{minipage}{7.9cm}
\epsfxsize 7.9cm \epsfbox{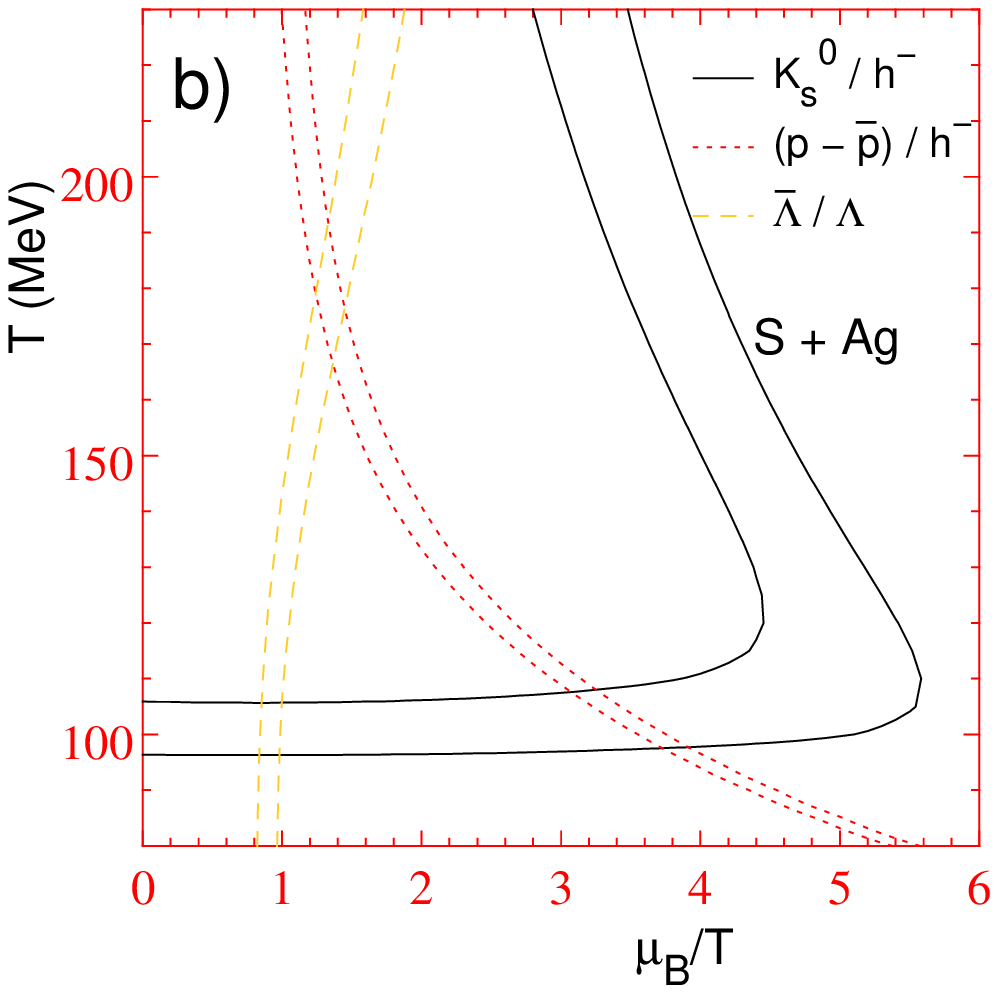}
\end{minipage}

\begin{minipage}{7.9cm}
\epsfxsize 7.9cm \epsfbox{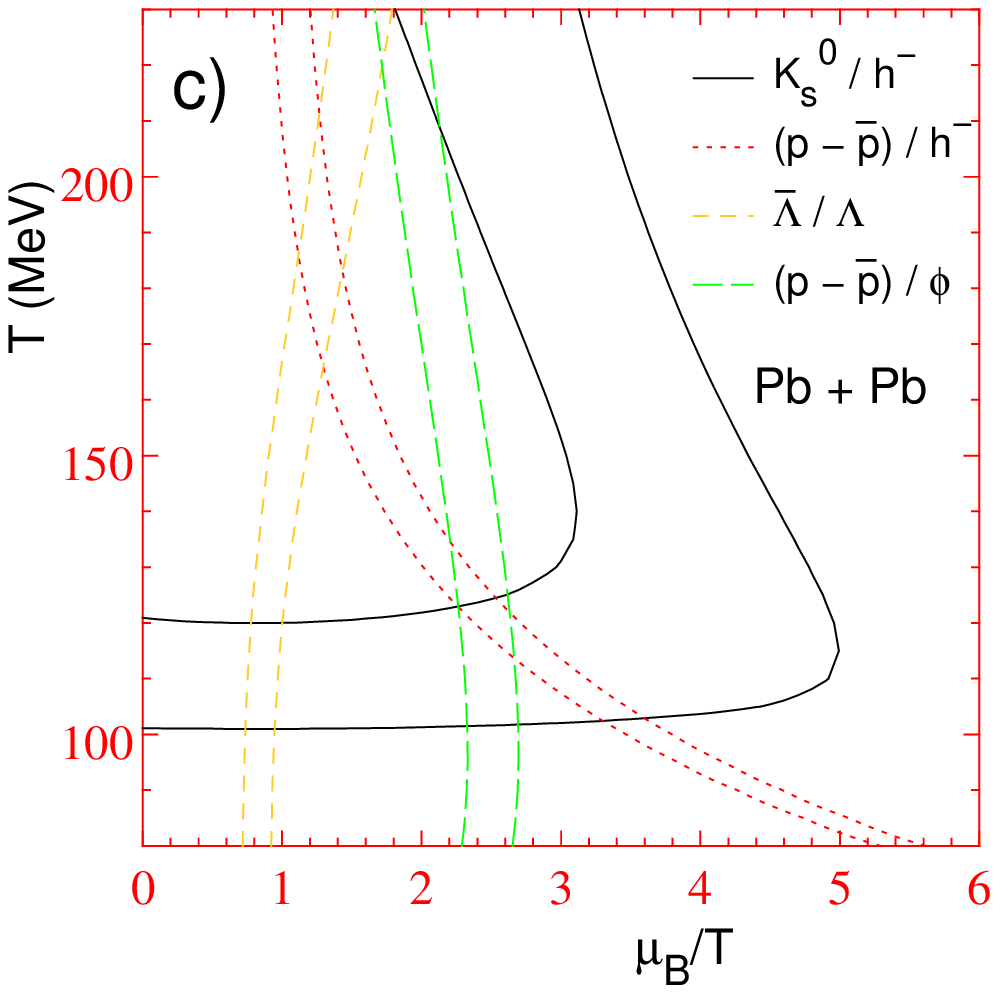}
\end{minipage}
\hfill
\begin{minipage}{7.9cm}
\end{minipage}
\end{figure}

In order to test this hypothesis we started the comparison of the experimental 
data with the hadron gas model by using first its fully equilibrated version, 
i.e. by setting $\gs=1$. We performed a graphical test similar to that 
made in Ref.~\cite{Redlich94} by plotting in the $T$-$\overline{\mu}_B/T$ plane the 
bands determined by the central values of some of the most relevant ratios 
of hadron yields and by their $1\sigma$ variations. As the overall 
multiplicities are very large, we set ${\sf A}^{-1} = 0$ in eq.~(\ref{seven}); 
in this grand-canonical limit the multiplicity ratios depend only on the 
intensive free parameters $T$ and $\overline{\mu}_B/T$ as the mean volume 
$\oV$ cancels out. This approximation turns out to be satisfactory for all 
the examined collisions as it is demonstrated below. In Fig.~\ref{gammasone} 
a--c such bands are shown for S+S, S+Ag and Pb+Pb data, respectively. 
There is no evident common crossing region for all bands, indicating 
absence of complete equilibrium even in central collisions of nuclei 
as heavy as Lead. It should be noted that a crossing region does exist
for Pb+Pb collisions with $T \simeq 120$ MeV and $\overline{\mu}_B/T 
\simeq$ 2.7, as long as only full phase space multiplicities are considered,
i.e. excluding the $\overline{\Lambda}/\Lambda$ band. 
However, these latter values imply an antibaryon/baryon ratio of less
than 10$^{-2}$ at primary level, which is unrealistically small with respect 
to the same ratio in S+S and S+Ag collisions. In fact the measured 
$\bar{\Lambda}/\Lambda$ band is quite far from the crossing point.

In order to confirm the previous finding we performed a least--squared 
fit to the data with $T$, $\oV$ and $\overline{\mu}_B/T$ as free 
parameters, using the canonical corrections and keeping $\gs = 1$ 
fixed. The details of the fitting procedure are described in Appendix C.
The results are shown in the first part of Table~\ref{chifit} and the 
fourth column of Table~\ref{comparison}. The $\chi^2/NDF$'s are about 
6$\div$8 with large discrepancies between fitted and measured values. 
Therefore we conclude that a full equilibrium version of the ideal 
hadron gas model fails to reproduce full phase space hadron multiplicities 
in central A+A collisions at CERN SPS energies. 

Our conclusion is in contradiction with the findings in 
Ref.~\cite{BraunMunzinger96}. This discrepancy can be explained by the fact 
that the analysis in Ref.~\cite{BraunMunzinger96} was performed by using 
particle ratios in various regions of rapidity and transverse momentum 
which, unlike in our analysis, requires additional dynamical input beyond
a simple statistical {\it ansatz}. Secondly, a proper statistical 
comparison between model predictions and data has not been performed
in Ref.~\cite{BraunMunzinger96}.

\subsection{Off--Equilibrium Model}

We tested the off--equilibrium version of the ideal hadron gas model 
repeating the fits with $\gs$ as a free parameter.
The results of these new fits are shown in Table~\ref{comparison} and 
in the lower half of Table~\ref{chifit} while the comparison between 
fitted and measured multiplicities (or ratios) is shown in 
Table~\ref{comparison}. There is a good agreement for all particles 
with some exceptions such as antiprotons in both S+S and S+Ag collisions. 
The $\chi^2/NDF$ for S+S and S+Ag collisions is about 3.5, thus significantly 
lower than for the full--equilibrium fit. The $\chi^2/NDF$ for central 
Pb+Pb collisions is close to 1.

The obtained temperatures and baryon chemical potentials are quite 
compatible with a common value and so are the $\gs$ values 
which turn out to be definitely less than 1 in all the three examined 
collisions. Also quoted in Table~\ref{chifit} are the obtained chemical 
potentials $\overline{\mu}_S$ and $\overline{\mu}_Q$ and the range of 
variation of the matrix ${\sf A}^{-1}$ elements; their smallness bears out 
the saddle--point approximation used in this analysis and the previously 
described graphical test by confirming the proximity to the grand-canonical 
regime.

\subsection{Discussion of the Analysis}

The dependence of the fitted parameters on the hadron mass spectrum
cut--off has been checked by repeating the fit with lower cut--off values
and found to be negligible.

The validity of the approximated averaged formula in eq.~(\ref{seven}) 
in the presence of participant fluctuations has been checked by calculating 
its first order corrections according to the formulae quoted in 
Appendix B with the assumption of moderate Gaussian fluctuations of 
baryon number $B$, electric charge $Q$ and volume $V$. 
The corrections have been estimated repeatedly by random variations of the 
correlations, assumed to be positive, between $B$, $Q$ and $V$ in order
to find out a maximum value. 
Particles undergoing the most significant variations with respect to the
average yield are baryons whose production increases mainly owing to terms 
proportional to $(\delta V/\oV)$ where $\delta\oV$ is the volume dispersion. 
Such variations turn out to be almost constant for the three kinds of 
collisions as a function of the relative dispersions of $B$, $Q$ and $V$. 
They are within 5\% with no volume fluctuation even for $\delta B/\oB$ and 
$\delta Q/\oQ$ of 30\% but can raise to more than 30\% if also a volume 
dispersion $\delta V/\oV = 30$\% is included.

As follows from the previous discussion (see Sect.~2) and the data 
presented in Table~\ref{comparison}, the relative width of baryon number 
distribution is expected to be around 6\% and 2\% for central S+S and 
Pb+Pb collisions, respectively. For these participant fluctuations 
the corrections to baryon yields range from about 
2.5\% (1\%) for $\delta V/\oV 
= 10$\% up to 17\% (13\%) for $\delta V/\oV = 30$\% in S+S (Pb+Pb) collisions.
It is worth remarking that the $B$, $Q$ and $V$ fluctuations would mainly 
influence the determination of the baryochemical potential in the fit as 
antibaryons and other particles production are much less affected by 
them compared with baryons. 
 
Whilst S+S and Pb+Pb collisions involve two identical nuclei, S+Ag is
an asymmetric nuclear collision and the use of the eq.~(\ref{eight}) 
to determine strange and electrical chemical potentials may be not appropriate.
According to eq.~(\ref{eight}) the average ratio $Q/B$ of participant
nucleons in S+Ag collisions is about 0.45 whereas it is actually likely
to be somewhat higher owing to the fact that Sulphur ($Z/A$=0.5) is the smaller
nucleus. In order to prove that our results are independent of the previous 
assumption, we repeated the fit by varying the charge/baryon number ratio 
and taking the extreme case $Q/B$=0.5. We found $T=177.5\pm7.9$ MeV, 
$\oV T^3 \exp(-0.7 {\rm GeV}/T) = 6.37\pm0.47$, $\gs = 0.707\pm0.063$,  
$\overline{\mu}_{B}/T = 1.325\pm0.081$ with a $\chi^2/NDF = 7.0/2$, a fit
result not significantly different from that quoted in Table~\ref{chifit}. 

A further technical problem in the $\chi^2$ fit is concerned with data 
redundancy. Whenever two or more data points are in a relationship that 
is not dependent on any free parameter, then the $\chi^2$ significance 
might be questionable as the effective number of degrees of freedom is 
overestimated.
This is the case for S+S collisions where the (approximate) relation 
$\langle {\rm K}^0_s \rangle  = (\langle {\rm K}^+\rangle  + 
\langle{\rm K}^-\rangle)/2$, owing to isospin symmetry and independent 
of model parameters, links three data points. However, it can be shown 
(see Appendix D), that such redundancy does not affect the extraction
of thermal parameters nor their errors; its main effect is a lowering
of the $\chi^2/NDF$. 

Our analysis is done in the ideal pointlike hadron gas framework. 
Since the extracted temperatures are very high, corrections due to 
particle repulsion in principle should be considered. However, most 
corrections 
proposed in literature leave particle ratios unchanged \cite{Sollfrank97b}
so that the parameters $T$, $\gs$ and chemical potentials are 
unaffected. On the other hand, our volume parameter $V$ is the 
pointlike particle volume and should not be confused with the actual volume. 
In a recent analysis \cite{Yen97} different particle ratios were compared
with the full--equilibrium hadron gas predictions by introducing different 
hard--core radii for pions and other hadrons leading to an effective pion 
chemical potential. Such a procedure restores the agreement of the pion 
abundance which is always underestimated using thermal parameters extracted 
from the remaining hadrons \cite{Sollfrank94} in a full--equilibrium model. 
However, the price to be payed is the introduction of additional, 
particle-species dependent, parameters.  

Our numerical results are quite different from a previous analysis of
S+S collisions \cite{Sollfrank94} which found $T= 205\pm39$ MeV, 
$\gs = 0.95\pm 0.20$ and $\mu_B/T = 1.30\pm 0.17$ by excluding negatively 
charged hadrons from the analysed data set. 
The fact that $\chi^2/NDF > 3$ in S+S collisions is expected to produce
fluctuations larger than 1$\sigma$ in the fitted parameters if some data
points are excluded in turn from the fit. In fact, by excluding negatives 
we found $T= 200\pm10$ MeV, $\gs = 0.99\pm0.11$ and $\mu_B/T=1.35\pm0.09$, 
quite in agreement with Ref.~\cite{Sollfrank94}.
The remaining difference can be explained by the use of an updated data sample
and particularly by the inclusion, in our analysis, of resonance widths; 
this especially enhances the $\rho$ meson production and, consequently, the 
pion production improving the agreement with the measured negatively charged 
hadron multiplicity.
 
Another recent analysis \cite{Letessier97} of hadron abundances in 
Pb+Pb collisions found $\gs = 0.9\pm0.09$ but with the use of only 
strange baryons ratios in a limited phase space region and omitting the 
$\phi$ multiplicity.
  
\section{Discussion and Conclusions}

\begin{figure}
\caption[]{Intensive thermal fit parameters as a function of system
size. Also plotted $T$ and $\gs$ fitted in \pp collisions at $\sqrt{s}=19.5$
GeV \cite{Becattini97a}.
\label{summary}} 

\begin{minipage}{15.0cm}
\epsfxsize 15.0cm \epsfbox{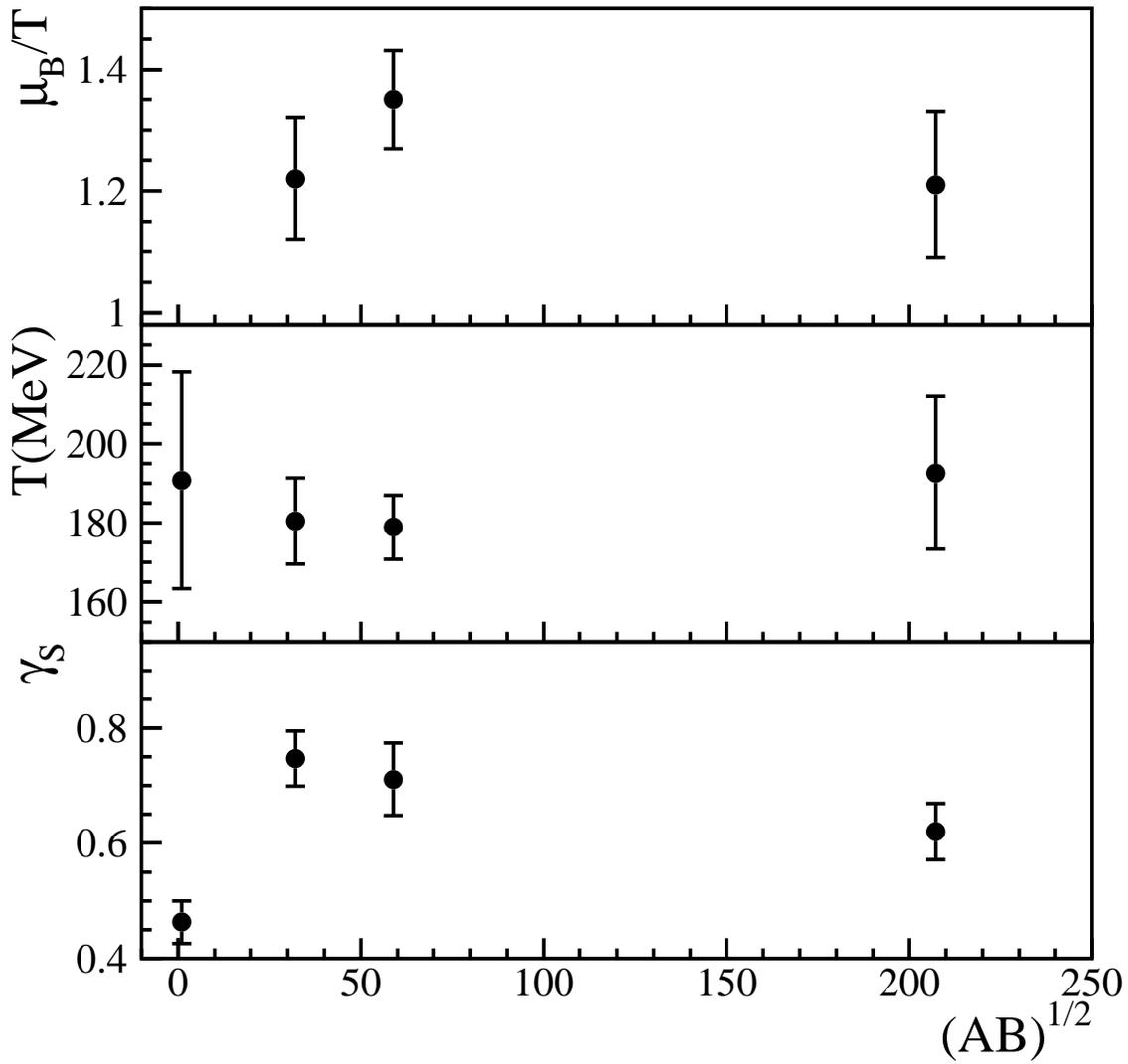}
\end{minipage}
\end{figure}

The data on hadron multiplicities in central A+A collisions at SPS energies
can not be reproduced by an ideal hadron gas assuming complete chemical 
equilibrium. This statement is also true for the central collisions of the 
heaviest nuclei, Pb+Pb.
The hadron gas model supplemented with partial strangeness saturation 
agrees significantly better with the data as the $\chi^2/NDF$ reduces from
about 7 to about 3 for S+S and S+Ag collisions and about 1 for Pb+Pb collisions.\\ 
The behaviour of the intensive parameters obtained within the off--equilibrium 
model is shown as a function of the system size in Fig.~\ref{summary} where
the result for the \pp data at similar collision energy \cite{Becattini97a} 
is also included. 

The lack of complete strangeness equilibrium at hadron level for 
central A+A collisions can not be interpreted as an effect of the choice 
of weights $w(\QG_1^0,\ldots,\QG_N^0)$ described in Sect.~2 which is crucial 
to reduce the number of free parameters. Since $\gs$ turns out to be 
$<1$, it might be argued indeed that if the hadron gas fireballs were small 
enough and all with zero strangeness (so that the weights 
$w(\QG_1^0,\ldots,\QG_N^0)$ would be no longer those chosen in Sect.~2), a 
suitable canonical suppression \cite{Becattini96a,Becattini97a} could be 
generated without the need of $\gs$ and a hadron gas in full chemical 
equilibrium would be recovered. Nevertheless this mechanism would have no 
effect on the yield of $\phi$ meson which is completely neutral, 
thus not suppressed by quantum number conservation at hadron level and 
having no known feeding from heavier light--flavoured resonances. 
Therefore, the measurement of $\phi$ production in Pb+Pb collisions 
establishes the necessity of a significant strangeness suppression at 
hadronic level independently of the validity of the assumed fireball 
quantum configurations occurrence probabilities.

An important conclusion can be drawn from the resulting chemical freeze--out
temperature, which seems to be independent of the system size and it
is, within errors, much the same as that in \ee, \pp and \ppb collisions 
\cite{Becattini96b,Becattini97a,Becattini97b}. This seems to indicate 
that the chemical freeze--out occurs close to the hadronization point and 
that the same mechanism of statistical hadronic phase space filling at 
critical parameters of the prehadronic matter invoked as a natural explanation 
of elementary collisions results \cite{Becattini97a,Becattini97b} also
holds for heavy ion collisions. Moreover, the similar $\mu_B/T$ values for all 
studied A+A collisions suggest a common hadronization nuclear density. 

As the fit is not perfect in S+S and S+Ag collisions even in the 
off--equilibrium model, one may speculate that the small deviations from 
model predictions are due to secondary inelastic interactions between hadrons 
following the hadronization stage. While thought to be absent in \pp collisions, 
they may likely occur in A+A collisions where they can destroy the statistical 
character of the hadronization process. In fact, as inelastic cross sections 
for different processes are significantly different, hadron rescatterings 
may lead to decoupling of different particle species at different temperatures, 
thus affecting the single temperature fit. 
This mechanism is particularly well--suited to explain the observed deviation 
of antiprotons which may quickly annihilate in the baryon--dense medium 
formed in an A+A collision. 
On the other hand it should be mentioned that the observed small deviations 
from the off--equilibrium version of the hadron gas model could simply stem 
from errors in the extrapolation procedures, from participant and volume 
fluctuations (as described in Sect.~3) and from the choice of weights 
$w(\QG_1^0,\ldots,\QG_N^0)$ (see Sect.~2).

Whilst the temperature is constant, the strangeness suppression factor 
increases from about 0.45 for \pp interactions to about 0.7 for central S+S 
collisions at comparable nucleon--nucleon centre of mass energies. No further 
increase of the strangeness suppression factor is observed for central 
Pb+Pb collisions. This observation has two important consequences: 
firstly, a heavy ion collision is not the result of an incoherent sum of 
nucleon collisions as far as strangeness production is concerned. 
In fact, due to isospin symmetry, $\gs$ must be the same in \pp and \nn 
collisions at the same $\sqrt s$; strangeness production in \pn interactions
was measured to be the same as in \pp interactions \cite{Ole}, thus 
the neutron content of colliding nuclei cannot account for the increase
of $\gs$ with respect to \pp and \ppb collisions. Secondly, as 
canonical strangeness suppression was taken into account in extracting 
$\gs$ in Ref.~\cite{Becattini97a} and in the present analysis, the 
strangeness enhancement in heavy ion collisions cannot be fully 
attributed to the increased system size at hadron level. 

\begin{figure}
\caption[]{The strangeness suppression factor $\LS = 
2\langle {\rm s}\bar{\rm s} \rangle/ (\langle {\rm u}\bar{\rm u} \rangle + 
\langle {\rm d}\bar{\rm d} \rangle)$ in high energy collisions as a function 
of centre of mass energy (nucleon--nucleon centre of mass energy for heavy 
ion collisions) calculated within the off--equilibrium hadron gas model. 
For \ee, \pp and \ppb collisions the ratios have been calculated by 
using model parameters quoted in Ref.~\cite{Becattini97b}. For \ppb 
in order to estimate the possible influence of the annihilation process, 
we plotted in addition the $\lambda_{\rm S}$ value calculated by 
including initial valence quarks and antiquarks (lower points).
For \ee collisions the leading s quarks in ${\rm e}^+{\rm e}^- \rightarrow 
{\rm s} \bar{\rm s}$ have been subtracted to calculate $\LS$. 
\label{su}} 

\begin{minipage}{15.0cm}
\epsfxsize 15.0cm \epsfbox{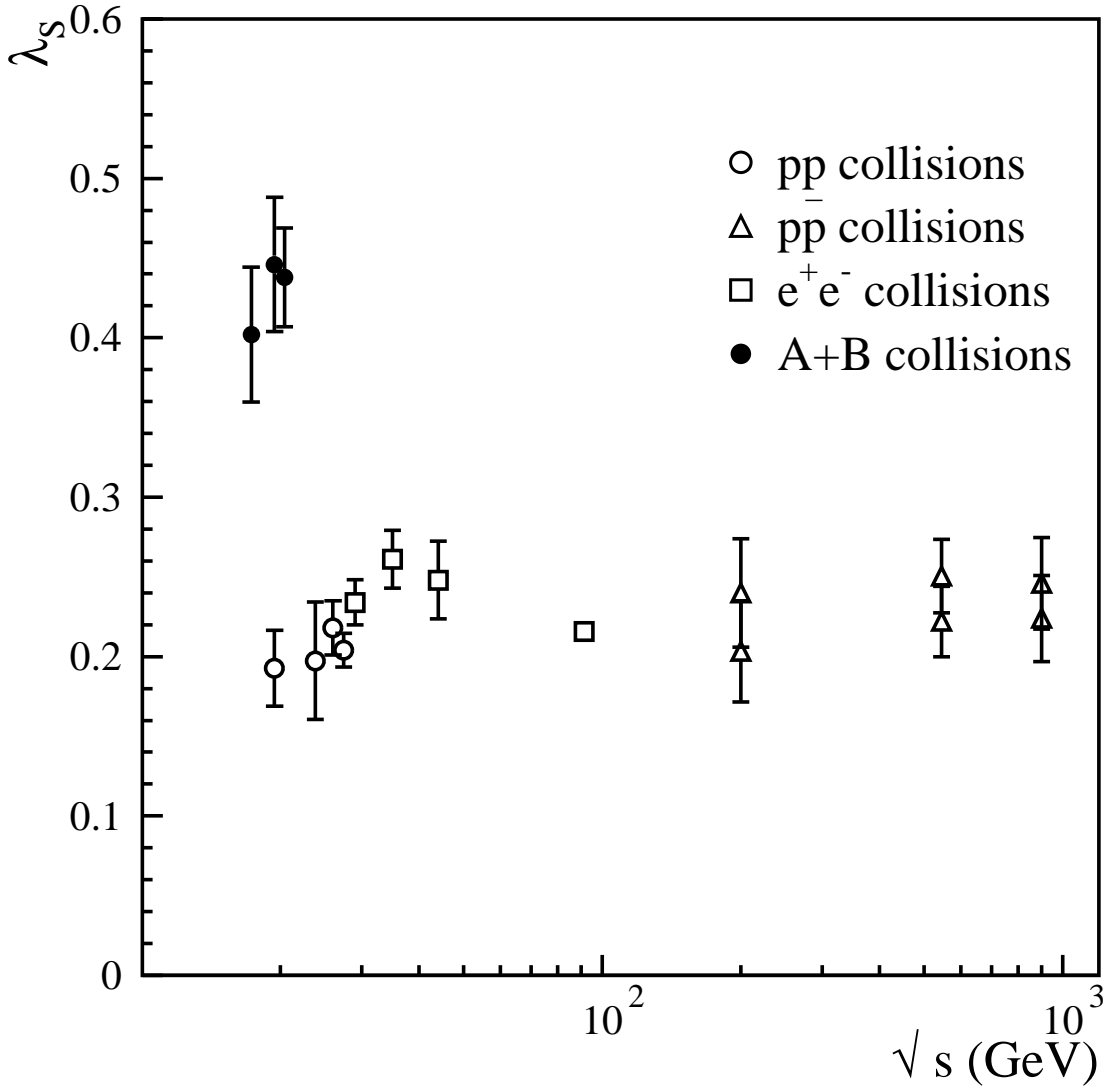}
\end{minipage}
\end{figure}

\begin{figure}
\caption[]{The strangeness suppression factor $\LS = 2\langle {\rm s}\bar{\rm s} 
\rangle/ (\langle {\rm u}\bar{\rm u} \rangle + \langle {\rm d}\bar{\rm d} 
\rangle)$ in a hadron gas at fixed $T$, $V$ and $\gs$ as a function of
the number of initial protons (baryon number equal to electric charge). 
\label{shg}} 

\begin{minipage}{15.0cm}
\epsfxsize 15.0cm \epsfbox{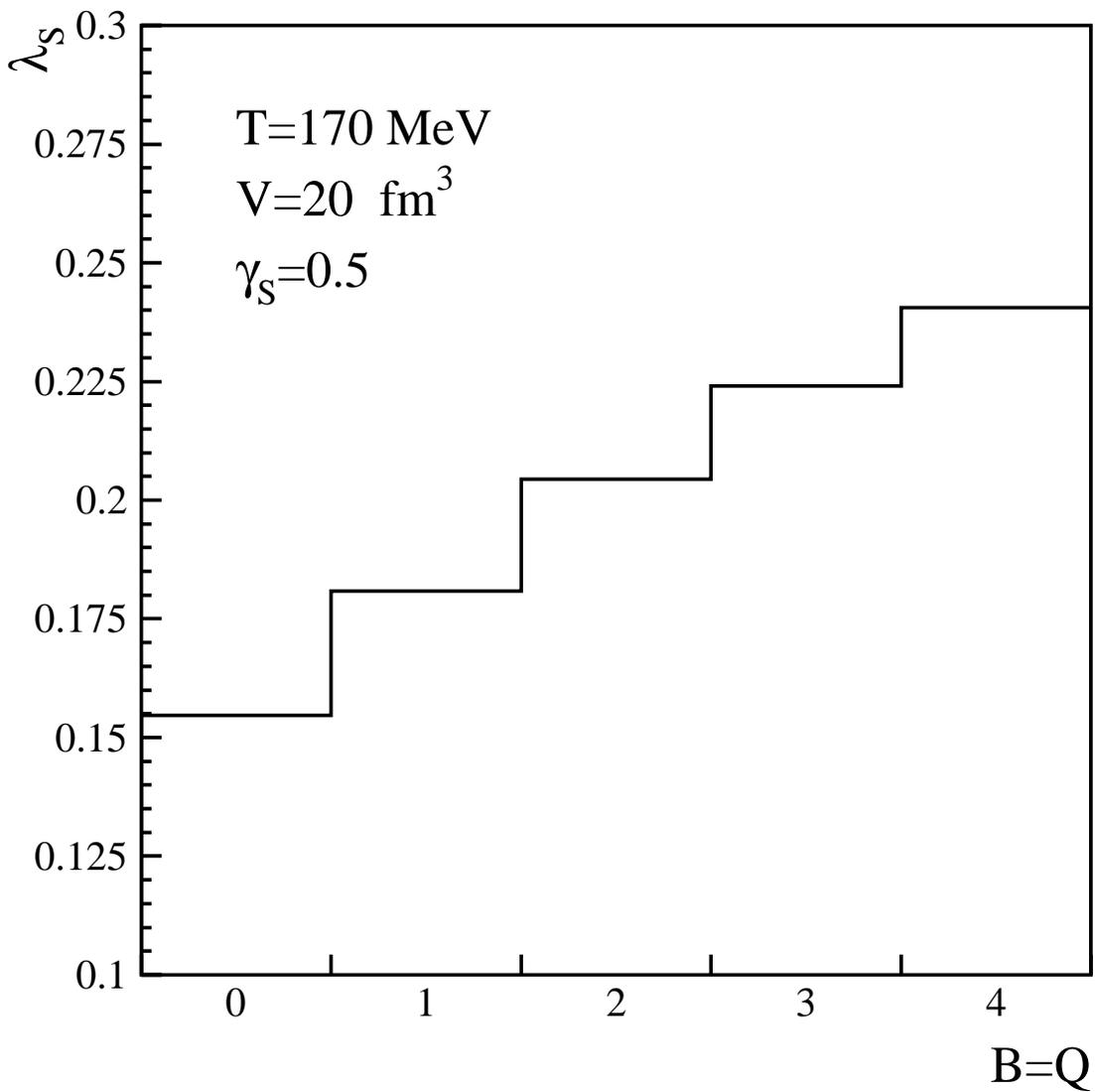}
\end{minipage}
\end{figure}

The relative production of strangeness has been intensively studied in 
elementary \cite{Ho:88} and nuclear collisions \cite{Bi:92,Ga:96}. 
It is usually expressed in terms of a strangeness suppression factor 
$\LS$ defined as: 

\begin{equation}
\LS = \frac {\langle {\rm s}\bar{\rm s} \rangle} 
{0.5(\langle {\rm u} \bar{\rm u} \rangle + \langle {\rm d} \bar{\rm d} \rangle)} \; ,
\label{ls}
\end{equation}
where $\langle{\rm s}\bar{\rm s}\rangle $, $\langle{\rm u}\bar{\rm u}\rangle$
and $\langle{\rm d}\bar{\rm d}\rangle$ are the mean multiplicities of newly
produced valence quark--antiquark pairs at primary hadron level, before 
resonance decays. Thus the initial colliding valence quarks are excluded
in calculating $\LS$. 
A major problem in the experimental determination of $\LS$ is to account 
for unmeasured hadron abundances. The statistical model used in this 
analysis is a useful tool for this purpose because it reproduces well 
all measured hadron abundances both in elementary and nuclear collisions, thus 
providing a reliable quark counting method. In Fig.~\ref{su} $\LS$ obtained 
by using model predictions for primary hadron multiplicities in \ee, \pp, \ppb
collisions with the parameters quoted in Ref.~\cite{Becattini97b} and for A+A 
collisions is shown. It should be mentioned that in \ee collisions the leading 
strange quarks in ${\rm e}^+{\rm e}^- \rightarrow {\rm s} \bar{\rm s}$ events 
have been subtracted from the numerator of eq.~(\ref{ls}) so that $\LS$ contains 
only valence quarks created during the hadronization process.
The $\LS$ values for elementary collisions are consistent with a constant value 
of about 0.2, even for very high energy \ppb collisions. The difference between 
$\gs$ in \ee ($\simeq 0.7$) compared to \pp, \ppb collisions ($0.46 \div 0.56$) 
as resulting from the off--equilibrium hadron gas model fit 
\cite{Becattini96b,Becattini97a} is mainly due to two effects. 
As far as \pp collisions are concerned, the presence of six initial u, d quarks 
to be hadronized along with those newly produced brings about a $\gs$ decrease.
In fact, for constant $T$, $V$ and $\gs$, a hadron gas with increasing baryon 
number and electric charge, i.e. increasing number of initial protons, has an 
increasing $\LS$ (see Fig.~\ref{shg}). The physical reason is the lower 
energy threshold for strange pair production in a baryon rich environment 
where the dominant process is via ${\rm N} + \pi \rightarrow \Lambda + {\rm K}$
while in the baryon free case it proceeds via kaon pair production.
Secondly, $T$ and $\gs$ are anticorrelated in \ee collisions. 
As the central fitted $T$ value is lower in \ee collisions 
in comparison with \pp and \ppb collisions \cite{Becattini97b}, the central
$\gs$ value is expected to be higher in \ee collisions to reproduce the 
measured strange hadron multiplicities. In fact, by repeating the same fit as in 
Ref.~\cite{Becattini96b} for \ee collisions at $\sqrt s=$ 29 and 91.2 GeV, 
and keeping fixed $T=170$ MeV instead of 160.3 MeV and 163.4 MeV respectively, 
$\gs$ turns out to be 0.69 and 0.62 instead of 0.72 and 0.67 respectively
with a slightly worse $\chi^2$.  

Our extracted value $\LS$ is in agreement with a previous estimate 
based on quark counting method quoted in Ref.~\cite{Wr:85} only for energies 
$\sqrt{s}<100$ GeV. On the other hand, the rise of $\LS$ in \ppb 
collisions claimed in Refs.~\cite{Wr:85,Wr:90} is not observed. The reason 
of this discrepancy is the fact that $\LS$ was estimated in 
Ref.~\cite{Wr:90} by using only K/$\pi$ ratio as experimental input and 
two parametrizations of hadron multiplicities \cite{AK,SS} which, unlike our
parametrization \cite{Becattini96b,Becattini97a}, do not satisfactorily 
reproduce all available measured multiplicities in \ppb collisions
\footnote{For instance, for \ppb collisions at $\sqrt s=$ 546 GeV, the 
parametrization in Ref.~\cite{AK} predicts a $\Lambda$/K$^0_s$ ratio of 0.49, 
taking the $\LS=0.28$ value quoted in Ref.~\cite{Wr:90}, whereas the 
experimental value is 0.24$\pm$0.05}.
 
To summarize, the common characteristics of elementary interactions seems 
to be the independence of $\LS$ on the collision energy, in the range 
examined in Refs.~\cite{Becattini96b,Becattini97a} and 
type of colliding particles. This universal behaviour is broken in central 
A+A collisions. The value of $\LS$ turns out to be about a factor 
two larger than the corresponding value for elementary interactions.
Our value for A+A collisions is consistent with previous estimates based on quark 
counting method \cite{Bi:92}. Note that in Ref.~\cite{Bi:92} $\LS$ 
was also estimated for p+A collisions and found to be consistent with that
of \pp collisions and independent of the size of the target nucleus.
This leads to the conclusion that no strangeness enhancement is observed 
in p+A collisions.

The saturation of $\gs$ and $\LS$ factors as a function of the 
colliding system size for central A+A collisions suggests that the strangeness 
enhancement with respect to elementary collisions may already occur in the 
prehadronic phase and that secondary hadron scatterings, expected to be much 
more abundant in Pb+Pb collisions, are of minor importance for strangeness 
production. The strangeness enhancement effect observed in central A+A collisions
and its independence of the colliding system size has been interpreted as due
to Quark--Gluon Plasma formation in the early stage of the collision 
\cite{Ga:96,Ga:97} already in S+S collisions and not only in central Pb+Pb 
collisions according to the interpretation of $J/\psi$ 
suppression \cite{Go:96,Bl:96}.

\section*{Acknowledgements}
J.S. acknowledges the support by the Bundesministerium f\"ur Bildung und
Forschung (BMBF) grant no. 06 BI 804 (5). We gratefully acknowledge
helpful discussions with J. Rafelski, H. Satz and R. Stock.

\newpage

\section*{Appendix} 
\begin{appendix}	
\section{Rapidity Distributions and Statistical Weights} 

In order to show that forward--backward peaked rapidity distributions 
for baryons and centrally peaked for antibaryons are not inconsistent with 
the assumption of statistical weights of eq.~(\ref{weight}) we consider a 
simple example of \pp collisions. Since the derived expression of average 
multiplicities in eqs.~(\ref{three},\ref{four}) does not depend either on 
the number of 
fireballs or their particular volumes $V_1,\ldots,V_N$, we consider a toy 
model with three fireballs with equal volume $V_f$ and sorted by 
boost velocities $\beta_1 > \beta_2 > \beta_3$. According to the statistical 
choice of weights $w(\QG_1^0,\ldots,\QG_N^0)$, owing to the equality of all
parameters $V_i$, $T_i$ and $\gsi$ of the fireballs, the probability of 
occurrence of baryon number configurations $\{1,0,1\}$, $\{0,1,1\}$ and 
$\{1,1,0\}$ is equal; the same holds for more complex and less probable 
sets of configurations such as $\{2,0,0\}$, $\{0,0,2\}$ and $\{0,2,0\}$.
Therefore, as far as average hadron multiplicities are concerned, nothing 
changes if one replaces $\{0,1,1\}$ and $\{1,1,0\}$ with $\{1,0,1\}$ and 
$\{0,2,0\}$ with $\{2,0,0\}$ in a half event sample and with $\{0,0,2\}$ in 
the remaining half. The hadron abundances do not vary and a strongly 
forward--backward peaked rapidity distribution for baryons can be obtained 
as the fireballs having a non--vanishing baryon number, are always those in 
the forward or backward directions.\\
In general this argument can be repeated for $N$ fireballs having an equal
rest frame volume and an arbitrary set of ordered boost velocities
$\beta_1 > \ldots > \beta_N$. In this case the weights in eq.~(\ref{weight})
are symmetric:

\begin{equation}
w(\QG_{\sigma(1)}^0,\ldots,\QG_{\sigma(N)}^0)=w(\QG_1^0,\ldots,\QG_N^0)
\end{equation}
for any permutation $\sigma$ of the integers $1,\ldots,N$. Therefore,
if $p(\QG_1^0,\ldots,\QG_N^0)$ are the actual weights, for eq.~(\ref{four})
to be valid, the condition to be fulfilled is:

\begin{equation}
 w[\QG_1^0,\ldots,\QG_N^0] = \frac{1}{N!} \sum_\sigma
 p(\QG_{\sigma(1)}^0,\ldots,\QG_{\sigma(N)}^0) \; ,
\end{equation}
where the square brackets mean that the set $[\QG_1^0,\ldots,\QG_N^0]$ is
a not--ordered one. This condition is weaker than a strict equality
between $w(\QG_1^0,\ldots,\QG_N^0)$ and $p(\QG_1^0,\ldots,\QG_N^0)$.

To summarize, the compatibility between the expression for hadron 
multiplicities (eqs.~(\ref{three},\ref{four})) and rapidity distributions 
can be achieved by choosing a model in which all fireballs have the same 
volume.
Their boost velocities and their total number are allowed to vary event 
by event and can be determined by using actual hadron spectra.
 
\section{Participant Nucleons and Volume Fluctuations} 

In this section we point out the conditions to be fulfilled for the 
replacement of eq.~(\ref{six}) with its averaged version eq.~(\ref{seven}) 
in the presence of
fluctuations of participant nucleons. In general, the variation of the number of
participants imply fluctuations of total baryon number $B$, electric charge 
$Q$ and also global volume $V$ of the colliding system. 
We assume that the associated distribution function $F(Q,B,V)$ is a Gaussian
and that the mean values $\oQ$ and $\oB$ are large, which is the case in
the examined collisions. If the latter condition is met, the sum over 
quantum vectors $\QGz$ in eq.~(\ref{six}) can be turned into an integration: 

\begin{eqnarray}
 \langle n_j \rangle &=& (2J_j+1) \frac{T}{2\pi^2} \, \sum_{l=1}^{\infty} 
 (\mp 1)^{l+1} \gs^{l s_j} \frac{m^2_j}{l} 
  {\rm K}_2(\frac{lm_j}{T}) 
  \nonumber \\
  && \quad\quad \times 
 \int \dint Q \dint B \dint V  \;F(Q,B,V) \;V \; 
 \E^{l \muvs \cdot \qj/T} \,\, \E^{-l^2 \qj {\sf{A}}^{-1} \qj/4} \; .
  \label{A1}
\end{eqnarray}
No more factor can be drawn out of the integral as the chemical potentials 
\muv depend on the integration variables because of the quantum number 
conservation constraint $\sum_j \qj \langle n_j \rangle =\QGz$ and the matrix 
{\sf A} is proportional to the volume (see eq.~(\ref{five})). 
Nevertheless, they can be expanded off the mean values $\oQ$, $\oB$ and 
$\oV$ up to first order, provided that the dispersions are not too large:

\begin{eqnarray}
 && \mbox{\muv} \simeq  \mbox{\muv}(\ovx) + {\sf J}_{\muvs} (\vx-\ovx) 
    \nonumber \\ 
 && {\sf A}^{-1} \simeq {\overline{{\sf A}^{-1}}} + 
    \frac{\partial{\sf A}^{-1}}{\partial V} (V-\oV) =
      {\overline{{\sf A}^{-1}}}\left(1-\frac{V-\oV}{\oV}\right)  \; .
\end{eqnarray}
where $\vx=(\oQ,\oB,\oV)$ and {\sf J} is the Jacobian matrix. Using the 
above expansions in eq.~(\ref{A1}) one obtains:

\begin{eqnarray}
 \langle n_j \rangle & = &\sum_{l=1}^{\infty} \langle n_j \rangle_l 
 \int \dint Q \dint B 
 \dint V \; \left(1+\frac{V-\oV}{\oV}\right) \;F(Q,B,V) \nonumber \\
 && \quad \quad \times 
\exp\left[l \qj \cdot {\sf J}_{\muvs} (\vx-\ovx)/T + l^2 \qj 
 {\overline{{\sf{A}}^{-1}}}\frac{(V-\oV)}{\oV} \qj/4 \right] \;,
\end{eqnarray}
where $\langle n_j \rangle_l$ is just the $l^{th}$ term of the series
in eq.~(\ref{seven}). The second term in the exponential is negligible 
if we are close to the grand-canonical regime and if the temperature 
is low enough to quickly suppress the terms of the series with high $l$: 
this condition is met for all hadrons if $T < 200$ MeV. Therefore:
 
\begin{equation} \label{X}
 \langle n_j \rangle \simeq \sum_{l=1}^{\infty} \langle n_j \rangle_l 
 \int \dint Q \dint B \dint V \; \left(1+\frac{V-\oV}{\oV}\right)\; F(Q,B,V) 
 \; \exp[l \qj \cdot {\sf J}_{\muvs} (\vx-\ovx)/T ] \; .
\end{equation}
If $F$ is a multivariate Gaussian:

\begin{equation}
   F(\vx) = \frac{1}{\sqrt{(2\pi)^3\det {\sf C}}} \exp[-(\vx- \ovx) 
   \cdot {\sf C}^{-1}(\vx-\ovx)/2] \; ,
\end{equation}
then the integral in eq.~(\ref{X}) can be solved analytically if the 
integration is extended to infinity. This is a satisfactory approximation 
if the dispersions are small compared to the mean values, which is
one of the basic requirements mentioned above. 

\begin{equation}\label{fourteen}
\langle n_j \rangle \simeq \sum_{l=1}^{\infty} \langle n_j \rangle_l
 \left[1+\frac{l}{T\oV}({\sf C} {\sf J}^T_{\muvs} \qj)_3\right]
 \exp \left[l^2 \qj \cdot {\sf J}_{\muvs} {\sf C} 
 {\sf J}^T_{\muvs} \qj /(2T^2)\right] \; .
\end{equation}
Thus, for the approximation (\ref{seven}) 
to be valid, it is necessary that 
${\sf C} {\sf J}^T_{\muvs}/(T\oV) \ll 1$ and ${\sf J}_{\muvs} {\sf C} 
{\sf J}^T_{\muvs}/T^2 \ll 1$.
The Jacobian matrix ${\sf J}_{\muvs}$ can be calculated by taking
the derivative of the quantum numbers conservation constraint:

\begin{equation}\label{XX}
 \frac{\partial}{\partial Q^0_i} \sum_j \qj n_j  = 
 \sum_j \qj \sum_{l=1}^{\infty} n_{jl} \frac{l}{T} 
   \sum_k q_j^k \frac{\partial \mu^k}{\partial Q^0_i} = {\bf e}_i \; ,
\end{equation}
where ${\bf e}_i$ is the $i^{th}$ unitary vector. If we define the 
matrix ${\sf B}$:

\begin{equation}
{\sf B}^k_i = \sum_j \sum_{l=1}^{\infty} l \; n_{jl} q_j^k q_j^i \; ,
\end{equation}
then the righthand equality in eq.~(\ref{XX}) can be inverted so to obtain
the derivatives of the chemical potentials:

\begin{equation}
  \frac{\partial \mu^k}{\partial Q^0_i} = T \left({\sf B}^{-1}\right)^k_i \; .
\end{equation}
It should be noted that the matrix ${\sf B}$ would be equal to 
$2{\sf A}$ if $\mbox{\muv}/T=0$. Since $\mbox{\muv}/T$ is generally 
${\cal O}(1)$ it turns out that ${\sf B}^{-1} = {\cal O}({\sf A}^{-1})$, 
hence it is expected to be much smaller than 1. To complete the Jacobian 
matrix ${\sf J}_{\muvs}$, we take the derivative of the quantum numbers 
conservation constraint with respect to $V$ for $V=\oV$, yielding:

\begin{equation}
\frac{{\sf B}}{T} \frac{\partial \mbox{\muv}}{\partial V} + \frac{\QGz}{\oV} -
 \frac{1}{4\oV} \sum_j \sum_{l=1}^{\infty} l^2 \qj {\overline{{\sf{A}}^{-1}}}
 \qj n_{jl} =0 \; .
\end{equation}
We use the Boltzmann limit for all hadrons in the last term, namely 
we keep only the first term of the series. By using this approximation, 
which is satisfactory if $T \simeq 170$ MeV, we conclude that the last term 
is $\simeq (1/4) {\sf B}^{-1} \QGz/ \oV$ which is much less than $\QGz/\oV$.
Therefore:

\begin{equation}
 \frac{\partial \mbox{\muv}}{\partial V} \simeq - T 
  {\sf B}^{-1}\frac{\QGz}{\oV} \; .
\end{equation}     
Finally, the Jacobian matrix ${\sf J}_{\muvs}$ turns out to be:

\begin{equation} 
   {\sf J}_{\muvs} \simeq T ({\cal O}({\sf A}^{-1}),{\cal O}
   ({\sf A}^{-1}),{\cal O}({\sf A}^{-1} \QGz/ \oV)) \; ,
\end{equation}
where each term is meant to be a column vector. This result can be used 
in conjunction with the eq.~(\ref{fourteen}) to establish the validity of the 
approximation (\ref{seven}). If ${\sf A}^{-1} \ll 1$ moderate fluctuations 
of $B$, $Q$, $V$ in comparison with the mean values are needed in order that 
${\sf C} {\sf J}^T_{\muvs}/T\oV \ll 1$ and ${\sf J}_{\muvs} 
{\sf C} {\sf J}^T_{\muvs}/T^2 \ll 1$. 

\section{Fitting Procedure}

We adopted a two--step fit procedure to also take into account the 
uncertainties on input parameters such as hadron masses, widths and 
branching ratios, which in principle can play a significant role in the 
test of the model. Firstly a $\chi^2$ with only experimental errors has 
been minimized and preliminary best--fit model parameters 
$T_0$, $\oV_0$, $\overline{\mu}_{B0}$ have been determined:
\begin{equation} \label{nine}
  \chi^2 = \frac{\sum_{i=1}^M (y_i^{\rm exp}-y_i^{\rm theo})^2}
    {\sigma_i^2} \; ,
\end{equation}
where the index $i$ runs over the $M$ data points. Keeping the preliminary 
model parameters fixed, the variations $\Delta y_i^{l{\rm theo}}$ of the 
multiplicities (or ratios) corresponding to the variations of the $l^{th}$ 
input parameter by one standard deviation have been calculated. Such 
variations have been considered as additional systematic uncertainties 
on the multiplicities and the following covariance matrix has been formed:

\begin{equation}
   C_{ij}^{\rm sys} = \sum_l \Delta y_i^l  \Delta y_j^l 
\end{equation}
to be added to the experimental covariance matrix $C^{\rm exp}$. Finally a 
new $\chi^2$ has been minimized:

\begin{equation}
   \chi^2 = \sum_{i,j=1}^M (y_i^{\rm exp}-y_i^{\rm theo})[(C^{\rm exp}+
      C^{\rm sys})^{-1}]_{ij}(y_j^{\rm exp}-y_j^{\rm theo}) 
\end{equation}
from which the best--fit estimates of the model parameters and their
errors have been obtained. Actually more than 130 among the most 
significant or worst known input parameters have been considered
and the corresponding 1$\sigma$ variations performed. This fit technique 
upgrades the one used by one of the authors in the analysis of thermal 
hadron production in \ee, \pp and \ppb collisions \cite{Becattini96a,
Becattini96b,Becattini97a} in that the off--diagonal elements of 
$C^{\rm sys}$ are also included.

\section{$\chi^2$ and Data Redundancy} 

We prove that the parameters fitted with a $\chi^2$ minimization and
their errors are not affected by the presence of redundant data. Let
$y_1 \ldots y_N$ be a set of experimental measurements among which 
$y_k \ldots y_N$ are measurements of the same variable. Let $y = f(x,\va)$ 
be the functional dependence to be tested where $\va$ is a set of parameters 
to be determined by means of a $\chi^2$ minimization:

\begin{equation}\label{bone}
  \chi^2 = \sum_{i=1}^N \frac {(y_i - f(x_i,\va))^2}{\sigma_i^2} \; .
\end{equation}
The eq.~(\ref{bone}) can be written also:

\begin{equation}
  \chi^2 = \sum_{i=1}^{k-1} \frac {(y_i - f(x_i,\va))^2}{\sigma_i^2}+
   \sum_{i=k}^N \frac {(y_i - \bar y + \bar y - f(x_k,\va))^2}{\sigma_i^2} \; ,
\end{equation}
where $\bar y$ is the weighted average of $y_k \ldots y_N$; all these
values correspond to the same abscissa $x_k$.
Hence:

\begin{equation}
  \chi^2 = \sum_{i=1}^{k-1} \frac {(y_i - f(x_i,\va))^2}{\sigma_i^2}+
       \sum_{i=k}^N \frac {(y_i - \bar y)^2}{\sigma_i^2} +
   \sum_{i=k}^N \frac {(\bar y - f(x_k,\va))^2}{\sigma_i^2} +
    2 \sum_{i=k}^N \frac {(y_i - \bar y)(\bar y - f(x_k,\va))}{\sigma_i^2} \; .
\end{equation}   
The second term in the above equation is simply the $\chi^2$ of the weighted
average while the third term can be written 
$(\bar y- f(x_k,\va))^2/\sigma_{\bar y}^2$, $\sigma_{\bar y}$ being the error
on the weighted average $\bar y$; the fourth term vanishes by definition of
weighted average. 
Therefore:

\begin{equation}
  \chi^2 = \chi^2_{WA} + \chi^2_{fit} \; ,
\end{equation}
where $\chi^2_{WA}$ is the $\chi^2$ of the weighted average and:

\begin{equation}
  \chi^2_{fit} = \sum_{i=1}^{k-1} \frac {(y_i - f(x_i,\va))^2}{\sigma_i^2}+
     \frac {(\bar y - f(x_k,\va))^2}{\sigma_{\bar y}^2} 
\end{equation}
is just the correct $\chi^2$ to minimize, for the $ N-k+1$ redundant points
have been replaced with their weighted average. Since $\chi^2_{WA}$ does
not depend on $\va$, the minimization of either $\chi^2$ or $\chi^2_{fit}$,
the latter being the correct one, leads to the same results. On the 
other hand, if $n={\rm dim}(\va)$ is the number of fitted parameters, 
the normalized $\chi^2_{fit}$ is:

\begin{equation}
  \chi^2_{fit} = \frac{\chi^2 - \chi^2_{WA}}{k-n}
\end{equation}  
instead of $\chi^2/(N-n)$. 

\end{appendix}	

\newpage


\newcommand{\IJMPA}[3]{{\it Int.~J.~Mod.~Phys.} {\bf A#1} (#2) #3}
\newcommand{\JPG}[3]{{\it J.~Phys. G} {\bf {#1}} (#2) #3}
\newcommand{\AP}[3]{{\it Ann.~Phys. (NY)} {\bf {#1}} (#2) #3}
\newcommand{\NPA}[3]{{\it Nucl.~Phys.} {\bf A{#1}} (#2) #3}
\newcommand{\NPB}[3]{{\it Nucl.~Phys.} {\bf B{#1}} (#2) #3}
\newcommand{\PLB}[3]{{\it Phys.~Lett.} {\bf {#1}B} (#2) #3}
\newcommand{\PRv}[3]{{\it Phys.~Rev.} {\bf {#1}} (#2) #3}
\newcommand{\PRC}[3]{{\it Phys.~Rev. C} {\bf {#1}} (#2) #3}
\newcommand{\PRD}[3]{{\it Phys.~Rev. D} {\bf {#1}} (#2) #3}
\newcommand{\PRL}[3]{{\it Phys.~Rev.~Lett.} {\bf {#1}} (#2) #3}
\newcommand{\PR}[3]{{\it Phys.~Rep.} {\bf {#1}} (#2) #3}
\newcommand{\ZPC}[3]{{\it Z.~Phys. C} {\bf {#1}} (#2) #3}
\newcommand{\ZPA}[3]{{\it Z.~Phys. A} {\bf {#1}} (#2) #3}
\newcommand{\JCP}[3]{{\it J.~Comp.~Phys.} {\bf {#1}} (#2) #3}
\newcommand{\HIP}[3]{{\it Heavy Ion Physics} {\bf {#1}} (#2) #3}

\begin{table*}[htb]
\begin{center}
 \caption[]{Comparison between fitted and measured hadron abundances and 
  ratios. All quoted multiplicities do not include feeding from weak 
  decays unless otherwise stated. Note: the $\chi^2$'s calculated by using
  values quoted below differ from those of Table~\ref{chifit} as the
  latter include contribution from uncertainties on input hadron parameters.
  \label{comparison}}

\vspace{0.5cm}
   \begin{tabular}{| c || c | c | c | c |}
    \hline\hline\noalign{\smallskip}
             &           &     & Fitted  & \\ 
  Hadron                  & Measured           & Fitted   & with $\gs$=1 & Reference \\ 
   \noalign{\smallskip} \hline\hline
\multicolumn{5}{|c|}{S+S collisions}\\
    \hline
   h$^-$  ($^{a}$)         & 98$\pm$3          &  92.63     &  82.04    &  \cite{Ro:94} \\ 
   K$^+$                   & 12.5$\pm$0.4      &  12.68     &  13.75    &  \cite{Ba:93} \\ 
   K$^-$                   & 6.9$\pm$0.4       &  7.611     &  7.785    &  \cite{Ba:93} \\ 
   K$^0_s$                 & 10.5$\pm$1.7      &  9.939     &  10.49    &  \cite{Al:94} \\ 
   $\Lambda$  ($^{b}$)     & 9.4$\pm$1.0       &  7.692     &  10.13    &  \cite{Al:94} \\ 
$\bar\Lambda$ ($^{b}$)     & 2.2$\pm$0.4       &  1.474     &  2.825    &  \cite{Al:94} \\ 
 p-$\bar{\rm p}$ ($^{c}$)  & 21.2$\pm$1.3      &  21.49     &  19.79    &  \cite{Ro:94} \\ 
 $\bar{\rm p}$  ($^{d}$)   & 1.15$\pm$0.4      &  2.092     &  2.314    &  \cite{Al:96} \\
    \hline
\multicolumn{5}{|c|}{S+Ag collisions}\\
    \hline
   h$^-$     ($^{a}$)      & 186$\pm$11    &  171.3     &  147.2        &  \cite{Ro:94}   \\ 
   K$^0_s$                 & 15.5$\pm$1.5  &  17.43     &  19.44        &  \cite{Al:94}   \\ 
   $\Lambda$ ($^{b}$)      & 15.2$\pm$1.2  &  13.99     &  17.44        &  \cite{Al:94}   \\ 
$\bar\Lambda$  ($^{b}$)    & 2.6$\pm$0.3   &  2.223     &  2.612        &  \cite{Al:94}   \\ 
 p-$\bar{\rm p}$ ($^{c}$)  & 43$\pm$3      &  43.44     &  39.18        &  \cite{Ro:94}   \\ 
 $\bar{\rm p}$  ($^{d}$)   & 2.0$\pm$0.8   &  3.381     &  2.401        &  \cite{Al:96}   \\
    \hline
\multicolumn{5}{|c|}{Pb+Pb collisions}\\
    \hline 
  Net baryon               & 372$\pm$10    & 375.7    & 372.6  &  \cite{Hu:97}   \\ 
   h$^-$         ($^{a}$)  & 680$\pm$50    & 650.2    & 638.5  &  \cite{Af:96}   \\ 
   K$^0_s$                 & 68$\pm$10     & 58.27    & 73.44  &  \cite{Af:96}   \\ 
   $\phi$                  & 5.4$\pm$0.7   & 5.759    & 5.648  &  \cite{Fr:97}   \\ 
 p-$\bar{\rm p}$ ($^{c}$)  & 155$\pm$20    & 155.3    & 147.8  &  \cite{Af:96}   \\ 
   K$^+$/K$^-$             & 1.8$\pm$0.1   & 1.652    & 1.700  &  \cite{Bo:97}   \\ 
$\bar\Lambda/\Lambda$      & 0.2$\pm$0.04  & 0.188    & 0.016  &  \cite{Bo:97}   \\
   \hline\noalign{\smallskip}
\multicolumn{5}{|l|}{$a$ - Defined as $\pi^- + {\rm K}^- + {\rm{\bar p}}$}\\
\multicolumn{5}{|l|}{$b$ - Includes feeding from $\Xi$}\\
\multicolumn{5}{|l|}{$c$ - Measured with the '+ - --' method, in this case limited}\\
\multicolumn{5}{|l|}{      rapidity acceptance (0.2-5.8) to exclude spectators}\\
\multicolumn{5}{|l|}{$d$ - Measured in a restricted rapidity interval and extrapolated}\\
\multicolumn{5}{|l|}{      by assuming that ${\rm{\bar p}}$ has the same rapidity distribution
                           as the $\bar\Lambda$}\\
\noalign{\smallskip}
   \hline\hline           
\end{tabular}
\end{center}
\end{table*}

\newpage

\begin{table*}[htb] 
\begin{center}
 \caption[]{Hadron gas model fitted parameters. The first set of parameters 
  has been obtained with a three--parameter fit by setting $\gs=1$. 
  The second set is the four--parameter fit result when only experimental 
  errors are used (first step of the fitting procedure) while the last set 
  is the final result including uncertainties on masses, widths and 
  branching ratios. Also quoted are the obtained chemical potentials, the 
  matrix ${\sf A}^{-1}$ elements and the $\chi^2$. The $\chi^2$ for S+S
  collisions within brackets is its corrected estimate accounting for
  kaons data redundancy (see text).\label{chifit}}

\vspace{0.5cm}
   \begin{tabular}{| c || c | c | c |}
        \hline\hline\noalign{\smallskip}
     Parameter                             &  S+S            & S+Ag            & Pb+Pb             \\ 
     \noalign{\smallskip}\hline\hline
      $T$ (MeV)                            &208.3$\pm$10.4   &179.9$\pm$7.8    &  125.4$\pm$4.6   \\
   $\oV T^3\exp[-0.7 {\rm GeV}/T]$         & 2.782$\pm$0.091 &4.91$\pm$0.30    &  13.1$\pm$1.1    \\  
      $\gs$ (fixed)                        &      1          &     1           &     1            \\  
      $\overline{\mu}_{B}/T$               & 1.145$\pm$0.066 &1.470$\pm$0.080  &  2.404$\pm$0.14  \\
      $\chi^2/$dof                         &   34.0/5        &  22.3/3         &    22.5/4        \\ \hline                      
      $T_0$ (MeV)                          & 182.4$\pm$9.2   & 181.8$\pm$6.9   &  192.6$\pm$8.1   \\
  $\oV_0T_0^3\exp[-0.7 {\rm GeV}/T_0]\!$   & 3.51$\pm$0.15   & 6.20$\pm$0.45   &  24.3$\pm$1.6    \\  
      $\gsz$                               & 0.732$\pm$0.038 & 0.727$\pm$0.057 &  0.616$\pm$0.043 \\  
  $\overline{\mu}_{B0}/T_0$                &1.248$\pm$0.074  & 1.365$\pm$0.072 &  1.207$\pm$0.071 \\
      $\chi^2_0/$dof                       &  17.1/4         &  7.74/2         &   3.99/3         \\ \hline
      $T$ (MeV)                            & 180.5$\pm$10.9  & 178.9$\pm$8.1   &  192.6$\pm$19.3  \\
  $\oV T^3\exp[-0.7 {\rm GeV}/T]$          & 3.48$\pm$0.16   & 6.29 $\pm$0.47  &  24.3$\pm2.2$    \\  
      $\gs$                                &0.747$\pm$0.048  & 0.711$\pm$0.063 &  0.620$\pm$0.049 \\  
  $\overline{\mu}_B/T$                     &1.22$\pm$0.10    & 1.350$\pm$0.081 &  1.21$\pm$0.12   \\
      $\chi^2/$dof                         &  12.4/4 (11.9/3)&   6.44/2        &  3.16/3          \\ 
  $\overline{\mu}_S/T$                     &  -0.320         &  -0.363         &  -0.372          \\
  $\overline{\mu}_Q/T$                     &  -0.00217       &  -0.0316        &  -0.0655         \\ 
      ${\sf A}^{-1}$           &$\!(-2.00\div 6.25)10^{-2}\!$&$\!(-1.10\div 3.54)10^{-2}\!$&$\!(-0.30\div 0.87)10^{-2}\!$ \\ 
   \hline\hline
\end{tabular}
\end{center}
\end{table*}

\newpage

\newpage

\newpage

\newpage

\end{document}